\documentclass{article}

\usepackage{times}
\usepackage{graphicx} 
\usepackage{subfigure} 

\usepackage{natbib}

\usepackage{algorithm}
\usepackage{algorithmic}

\usepackage{xspace}

\usepackage{amsmath}

\usepackage{hyperref}

\def\Ntot{\ensuremath{N_{\mathrm{tot}}}\xspace}
\def\Nbin{\ensuremath{N_{\mathrm{leaves}}}\xspace}
\def\vecx{\ensuremath{\mathbf{x}}\xspace}
\def\sign{\ensuremath{\mathrm{sign}}\xspace}
\newcommand{\binfunc}[1]{\ensuremath{\mathrm{leaf}_{\mathit{#1}}}\xspace}
\newcommand{\nodefunc}[1]{\ensuremath{\mathrm{node}_{\mathit{#1}}}\xspace}
\newcommand{\Nbinfunc}[1]{\ensuremath{N(\mathrm{leaf}_{\mathit{#1}})}\xspace}
\newcommand{\Vbinfunc}[1]{\ensuremath{V(\mathrm{leaf}_{\mathit{#1}})}\xspace}
\newcommand{\Ifunc}[2]{\ensuremath{\mathcal{I}(\mathit{#1}_{\mathit{#2}}})\xspace}

\title{Density Estimation Trees in High Energy Physics}
\author{Lucio Anderlini \\ Istituto Nazionale di Fisica Nucleare -- 
        Sesto Fiorentino, Firenze}
\date{January 30th, 2015}

\begin{document} 
\maketitle
%
%
%
%

\begin{abstract} 
  Density Estimation Trees can play an important role in exploratory data 
  analysis for multi-dimensional, multi-modal data models of large samples.
  I briefly discuss the algorithm, a self-optimization technique based on 
  kernel density estimation, and some applications in High Energy 
  Physics.
\end{abstract} 

\section{Introduction}
The usage of nonparametric density estimation techniques has seen a quick growth
in the latest years both in High Energy Physics (HEP) and in other fields of 
Science 
dealing with multi-variate data samples. Indeed, the improvement in the 
computing resources available for data analysis allows today to process a much
larger number of entries requiring more accurate statistical models.
Avoiding parametrization for the distribution with respect to one or more 
variables allows to enhance accuracy removing unphysical constraints on the 
shape of the distribution. The improvement becomes more evident when considering
the joint probability density function with respect to correlated variables,
for whose model a too large number of parameters would be required. 

Kernel Density Estimation (KDE) is a nonparametric density estimation technique 
based on the estimator
\begin{equation}
  \hat f_{\mathrm{KDE}}(\vecx) = \frac{1}{\Ntot}\sum_{i = 1}^{\Ntot} k(\vecx - \vecx_i),
\end{equation}
where $\vecx  = (x^{(1)}, x^{(2)}, ..., x^{(d)})$ is the vector of 
coordinates of the $d$-variate space $\mathcal S$
describing the data sample of \Ntot entries, $k$ is a normalized function
referred to as \emph{kernel}.
KDE is widely used in HEP \cite{Cranmer:2000du,Poluektov:2014rxa} including 
notable applications to the Higgs boson mass measurement by the ATLAS 
Collaboration \cite{Aad:2014aba}. The variables considered in the construction
of the data-model are the mass of the Higgs boson candidate and the response
of a \emph{Boosted Decision Tree} (BDT) algorithm used to \emph{classify} the 
data entries as \emph{Signal} or \emph{Background} candidates \cite{cart84}. 
This solution
allows to synthesize a set of variables, input of the BDT, into a single 
variable, the BDT response, which is modeled.
In principle, a multivariate data-model of the BDT-input variables may simplify
the analysis and result into a more powerful discrimination of signal and 
background. Though, the computational cost of traditional nonparametric 
data-model (histograms, KDE, ...) for the
sample used for the training of the BDT, including $\mathcal O(10^6)$ entries,
is prohibitive.

Data modelling, or density estimation, techniques based on decision trees are
discussed in the literature of statistics and computer vision communities 
\cite{ram2011density,provost2000well}, 
and with some optimization they are suitable for HEP as they 
can contribute to solve both classification and analysis-automation problems
in particular in the first, exploratory stages of data analysis.

In this paper I briefly describe the Density Estimation Tree (DET) algorithm, 
including an innovative and fast cross-validation technique based on KDE and 
consider few examples of successful usage of DETs in HEP.

\section{The algorithm}
A decision tree is an algorithm or a flowchart composed of internal \emph{nodes}
representing tests of a variable or of a property. Nodes are connected to
form \emph{branches}, each terminates into a \emph{leaf}, associated to a 
\emph{decision}.
Decision trees are extended to Density (or Probability) Estimation Trees when 
the \emph{decisions} are probability density estimations of the underlying 
probability density function of the tested variables.
Formally, the estimator is written as
\begin{equation}
  \hat f (\vecx) = \sum_{i = 1}^{\Nbin} \frac{1}{\Ntot} 
                   \frac{\Nbinfunc{i}}{\Vbinfunc{i}} 
                   \Ifunc{\vecx}{i},
\end{equation}
where \Nbin is the total number of leaves of the decision tree, \Nbinfunc{i}
the number of entries associated to the $i$-th leaf, and \Vbinfunc{i} is its
volume. If a generic data entry, defined by the input variables $\vecx$, 
would fall within the $i$-th leaf, then \vecx is said to be in the $i$-th leaf,
and the characteristic function of the $i$-th leaf,
\begin{equation}\label{eq:detestimator}
  \Ifunc{\vecx}{} = \left\{
    \begin{array}{ll} 
      1 & \mbox{if $\vecx \in \binfunc{i}$} \\
      0 & \mbox{if $\vecx \not\in \binfunc{i}$} \\
    \end{array}
    \right.,
\end{equation}
equals unity. By construction, all the characteristic functions associated to the
other leaves, are null. Namely,
\begin{equation}
  \vecx \in \binfunc{i} \quad\Rightarrow\quad \vecx \not\in \binfunc{j} 
  \quad\forall j : j \neq i.
\end{equation}

The training of the Density Estimation Tree is divided in three steps: 
\emph{tree growth}, \emph{pruning}, and \emph{cross-validation}. Once the 
tree is trained it can be evaluated using the simple estimator of Equation
\ref{eq:detestimator} or some evolution obtained through \emph{smearing} or
\emph{interpolation}. These steps are briefly discussed below.
\subsection{Tree growth}
As for other decision trees, the tree growth is based on the 
minimization of an estimator of the error. For DETs, the error is the 
Integrated Squared Error (ISE), defined as
\begin{equation}
  \mathcal R = \mathrm{ISE} (f, \hat f) = \int_{\mathcal S} 
                  (\hat f(\vecx) - f(\vecx))^2 
                 \mathrm d\vecx.
\end{equation}
It can be shown (see for example \cite{anderlini:2015} for a pedagogical discussion)
that, for large samples, the minimization of the ISE is equivalent to 
the minimization of 
\begin{equation}
  \mathcal R_{\mathrm{simple}} = -\sum_{i = 1}^{\Nbin}
    \left( \frac{\Nbinfunc{i}}{\Ntot}\right)^2 \frac{1}{\Vbinfunc{i}}.
\end{equation}
The tree is therefore grown by defining the replacement error 
\begin{equation}
  R(\binfunc{i}) = -\frac{\big(N(\binfunc{i})\big)^2}{\Ntot^2 \Vbinfunc{i}},
\end{equation}
and iteratively splitting each leaf $\ell$ to two sub-leaves $\ell_L$ and 
$\ell_R$ maximising the residual gain 
\begin{equation}
  G(\ell) = R(\ell) - R(\ell_L) - R(\ell_R).
\end{equation}
The growth is arrested, and the splitting avoided,
when some stop condition is matched. The most common
stop condition is $N(\ell_L) < N_{\mathrm{min}}$ or $N(\ell_R) < N_\mathrm{min}$;
but it can be OR-ed with some alternative requirement, for example on the 
widths of the leaves.

A more complex stop condition is obtained by defining a minimal leaf-width 
$t^{(m)}$ with respect to each dimension $m$.
Splitting by testing $x^{(m)}$ is forbidden if the width of one of the resulting
leaves is smaller than $t^{(m)}$.
When no splitting is allowed the branch growth is stopped.
This stop condition requires to issue the algorithm with a few more input 
parameters, the leaf-width thresholds, but is very powerful against
over-training. Besides, the determination of  reasonable leaf-widths is 
an easy task for most problems, once the expected resolution on each
variable is known.

\begin{figure}
  \centering
  \includegraphics[width=0.4\textwidth]{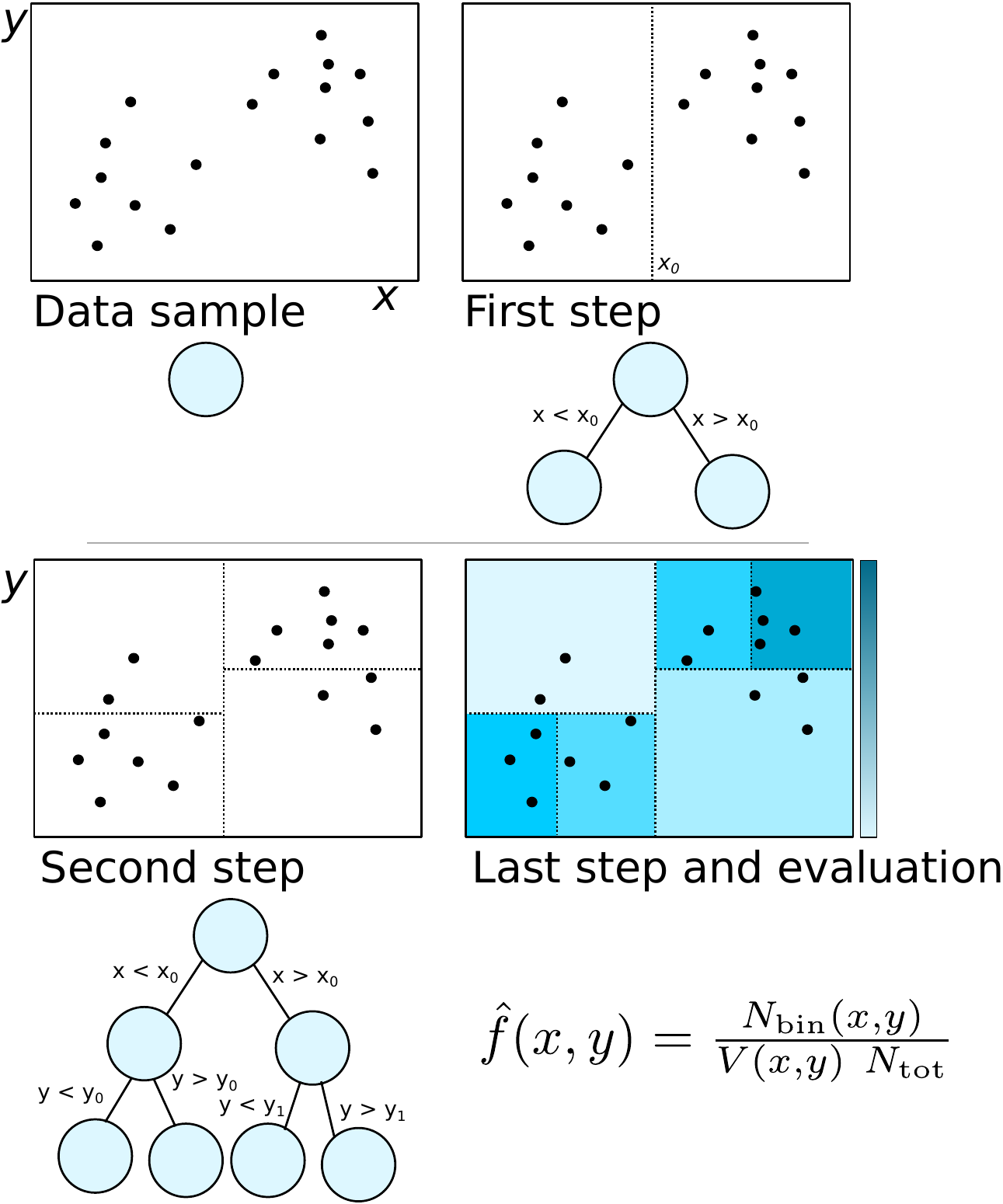}
  \caption{\label{fig:trainingexample}
    Simple example of training of a density estimation tree over a 
    two dimensional sample.
    }
\end{figure}

Figure \ref{fig:trainingexample} depicts a simple example of the training 
procedure on a two-dimensional real data-sample.

\subsection{Tree pruning}
DETs can be overtrained. Overtraining (or overfitting)
occurs when the statistical model obtained through the DET describes random 
noise or fluctuations instead of the underlying distribution.
The effect results in trees with isolated leaves with small volume and therefore
associated to large density estimations, surrounded by almost-empty leaves.
Overtraining can be reduced through \emph{pruning}, an \emph{a posteriori}
processing of the tree structure. The basic idea is to sort the nodes in terms
of the actual improvement they introduce in the statistical description of the 
data model.
Following a procedure common for classification and regression trees, the 
\emph{regularized error} is defined as 
\begin{equation}
  R_\alpha(\nodefunc{i}) = \sum_{j \in \mathrm{leaves\ of\ \nodefunc{i}}}
      \!\!\!\!\!\!\!\!\! R(\binfunc{j}) + \alpha C(\nodefunc{i}),
\end{equation}
where $\alpha$ is named \emph{regularization parameter}, and the index $j$
runs over the sub-nodes of \nodefunc{i} with no further sub-nodes (its leaves).
$C(\nodefunc{i})$ is the \emph{complexity function} of \binfunc{i}.

Several choices for the complexity function are possible. In the literature 
of classification and regression trees, a common definition is to set
$C(\nodefunc{i})$ to the number of terminal nodes (or leaves) attached to 
\nodefunc{i}. Such a complexity function provides a top-down simplification 
technique which is complementary to the stop condition. Unfortunately, 
in practice, the optimization through the pruning obtained with a 
number-of-leaves complexity function is ineffective against overtraining, if 
the stop condition is suboptimal.

An alternative cost function, based on the depth of the node in the tree 
development, provides a bottom-up pruning, which can be seen as an \emph{a 
posteriori} optimization of the stop condition.

\begin{figure}
  \centering
  \includegraphics[width=0.3\textwidth]{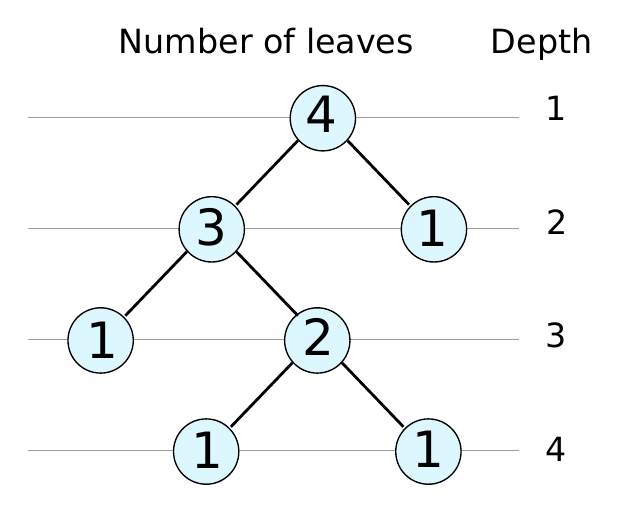}
  \caption{\label{fig:complexityFunction}
      Two examples of complexity function based on the number of leaves or 
      subtrees, or on the node depth.
    }
\end{figure}

An example of the two cost functions discussed is shown in Figure 
\ref{fig:complexityFunction}.

If $R_\alpha(\nodefunc{i}) > R(\nodefunc{i})$ the splitting of the $i$-th
node is pruned, and its sub-nodes merged into a unique leaf.
Each node is therefore associated to a threshold value of the regularization
parameter, so that if $\alpha$ is larger than the threshold $\alpha_i$, then
the $i$-th node is pruned. Namely,
\begin{equation}
  \alpha_i = \frac{1}{C(\nodefunc{i})} \left( R(\nodefunc{i}) - 
  \!\!\!\!\!\!
  \!\!\!\!\!\!
  \sum_{j \in \mathrm{leaves\ of\ \nodefunc{i}}}
  \!\!\!\!\!\!
  \!\!\!\!\!\!
  R(\binfunc{j}) \right)
  .
\end{equation}

The quality of the estimator $Q(\alpha)$, defined and discussed below, can
then be evaluated per each threshold value of the regularization parameter.
The optimal pruning is obtained for 
\begin{equation}
  \alpha = \alpha_{\mathrm{best}} \quad : 
  \quad Q(\alpha_{\mathrm{best}}) = \max_{\alpha \in \{\alpha_i\}_i} Q (\alpha).
\end{equation}

\subsection{Cross-validation}
The determination of the optimal regularization parameter is named 
\emph{cross-validation}, and many different techniques are possible, depending
on the choice of the quality function.

A common cross-validation technique for classification and regression trees
is the \emph{Leave-One-Out} (LOO) cross-validation and consists in the estimation
of the underlying probability distribution through a resampling of the original
dataset. For each data entry $i$, a sample containing all the entries but $i$
is used to train a DET. The ISE is redefined as
\begin{equation}
  R_{\mathrm{LOO}}(\alpha) = 
    \int_{\mathcal S} \left(\hat f^{\alpha}(\vecx)\right)^2\mathrm d\vecx
    -\frac{2}{\Ntot} \sum_{i=1}^{\Ntot} \hat f_{\mathrm{not}\ i}^\alpha(\vecx_i)
    ,
\end{equation}
where $\hat f^\alpha(\vecx)$ is the probability density estimation obtained 
with a tree pruned with regularization parameter $\alpha$, and 
$\hat f_{\mathrm{not}\ i}^\alpha(\vecx)$  is the analogous estimator obtained 
from a dataset obtained removing the $i$-th entry form the original sample.
The quality function is 
\begin{equation}
  Q(\alpha) = - R_{\mathrm{LOO}}(\alpha).
\end{equation}

The application of the LOO cross-validation is very slow and requires to 
build one decision tree per entry. When considering the application of 
DETs to large samples, containing for example one million of entries, the 
construction of a million of decision trees and their evaluation per one 
million of threshold regularization constants becomes unreasonable.

A much faster cross-validation is obtained comparing the estimation obtained 
with the DET with a triangular-kernel density estimation 
\begin{align}\nonumber
  f_k & (\vecx) = \frac{1}{\Ntot}\times \\ 
      & \times \sum_{i=1}^{\Ntot} 
      \prod_{k=1}^{d}\left(1-\left|\frac{\vecx - \vecx_i}{h_k}\right|\right) 
         \,\theta\!\left(1-\left|\frac{\vecx - \vecx_i}{h_k}\right|\right),
\end{align}
where $\theta(x)$ is the Heaviside step function, $k$ runs over the $d$
dimensions of the coordinate space $\mathcal S$, and $h_k$ is the 
kernel bandwidth with respect to the variable $x^{(k)}$.

The quality function is
\begin{equation}
  Q^{ker}(\alpha) = - \int_{\mathcal S} 
    (f_\alpha(x)^2 - f_k(x)^2)^2 \mathrm d\vecx.
\end{equation}
The choice of a triangular kernel allows to analytically solve the integral
writing that
\begin{equation}
  Q^{ker}(\alpha) = \frac{1}{\Ntot^2}\sum_{j = 1}^{\Nbin} 
    \frac{N(\binfunc{j}^\alpha)}{V(\binfunc{j}^\alpha)}
    (2 \mathcal N_j - N(\binfunc{j}^\alpha))
    + \mathrm{const},
\end{equation}
where $\binfunc{j}^\alpha$ represents the $j$-th leaf of the DET pruned with 
regularization constant $\alpha$, and 
\begin{align}\nonumber
  \mathcal N_j = & \sum_{i=1}^{\Ntot}\sum_{k=1}^{d}\mathcal I_{jk}(\vecx_i;h_k)  
               =  \sum_{i=1}^{\Ntot}  \sum_{k=1}^{d} 
        \Bigg[ u_{ij}^{(k)}-\ell_{ij}^{(k)} + \\
  \nonumber
    - & \frac{(u_{ij}^{(k)} - x_i^{(k)})^2}{2h_k}
        \sign\left(u_{ij}^{(k)}-x_i^{(k)}\right) + \\ 
    + & \frac{(\ell_{ij}^{(k)} - x_i^{(k)})^2}{2h_k}
        \sign\left(x_i^{(k)}-\ell_{ij}^{k)}\right)\Bigg].
   \label{eq:nj}
\end{align}
with $\sign(x) = 2\theta(x) - 1$, and 
\begin{equation}\label{eq:uijlij}
  \left\{
  \begin{array}{l}
    u_{ij}^{(k)}=\min\left(x_{\max}^{(k)}(\binfunc{j}), x_i^{(k)} + h_k\right)\\
    \ell_{ij}^{(k)}=\max\left(x_{\min}^{(k)}(\binfunc{j}),x_i^{(k)}-h_k\right).
  \end{array}
  \right.
\end{equation}
In Equation \ref{eq:uijlij}, $x_{\max}^{(k)}(\binfunc{j})$ 
and $x_{\min}^{(k)}(\binfunc{j})$ represent
the upper and lower boundaries of the $j$-th leaf, respectively.

An interesting aspect of this technique is that a large part of the 
computational cost is hidden in the definition of $\mathcal N_j$ which 
does not depend on $\alpha$, and therefore can be calculated only once per 
node, \emph{de facto} reducing the computational complexity by a factor
$\Ntot \times \Nbin$.

\subsection{DET Evaluation: smearing and interpolation}
One of the major limitations of DETs is the existence of sharp boundaries which
are unphysical. Besides, a small variation of the position of a boundary can
lead to a large variation in the final result, when using DETs for data 
modelling. Two families of solutions are discussed here: smearing and 
linear interpolation. The former can be seen as a convolution of the density
estimator with a resolution function. The effect is that sharp boundaries 
disappear and residual overtraining is cured, but as long as the resolution 
function has a fixed width, the adaptability of the DET algorithms 
is partially lost: resolution will never be smaller than the smearing function 
width. 

An alternative technique is interpolation, assuming some behaviour 
(usually linear) of the density estimator between the middle points of each 
leaf. The density estimation at the center of each leaf is assumed to be 
accurate, therefore overtraining is not cured, and may lead to catastrophic 
density estimations. Interpolation is treated here only marginally.
It is not very robust, and it is hardly scalable to more than two dimensions.
Still, it may represent a useful smoothing tool for samples composed of
contributions with resolutions spanning a large interval, for which 
adaptability is crucial.

\subsubsection{Smearing}
The smeared version of the density estimator can be written as
\begin{equation}
  \hat f_s (\vecx) = \int_{\mathcal S} \hat f(
    \mathbf{z}) w\left(\frac{\vecx - 
    \mathbf{z}}{h_k}\right)\mathrm d\mathbf{x},
\end{equation}
where $w(\vecx)$ is the \emph{resolution function}. 
Using a triangular resolution function $w(t) = (1-|t|)\,\theta(1-|t|)$, 
\begin{equation}
  \hat f_s (\vecx) = \sum_{j = 1}^{\Nbin} \prod_{k=1}^{d}
    \mathcal I_{jk}(\vecx; h_k),
\end{equation}
where $\mathcal I_{jk}(\vecx; h_k)$ was defined in Equation \ref{eq:nj}.

Note that the evaluation of the estimator does not require a loop on the 
entries, factorized within $\mathcal I_{jk}$.

\subsubsection{Interpolation}
As mentioned above, the discussion of interpolation is restrained to 
two-dimensional problems. The basic idea of linear interpolation is to associate
each $\vecx \in \mathcal S$ to the three leaf centers  defining the smallest
triangle inscribing $\vecx$ (step named \emph{padding} or \emph{tessellation}). 
Using the positions of the leaf centers, and the corresponding values of the 
density estimator as coordinates, it is possible to define a unique plane.
The plane can then be ``read'' associating to each $\vecx \in \mathcal S$ 
a different density estimation.
The key aspect of the algorithm is \emph{padding}. Padding techniques are 
discussed for example in \cite{deBerg:2008}. The algorithm used in the 
examples below is based on Delaunay tessellation as implemented in the ROOT
libraries \cite{ROOT:1971}.
Extensions to more than two dimensions are possible, but non trivial and 
computationally expensive. Instead of triangles, one should consider 
hyper-volumes defined by $(d+1)$ leaf centers, where $d$ is the number of 
dimensions.
Moving to parabolic interpolation is also reasonable,
but the tessellation problem for $(d+2)$ volumes is less treated in the 
literature, requiring further development.

\section{Timing and computational cost}
The discussion of the performance of the algorithm is based on an a 
single-core C++ implementation. Many-core tree growth, with each core 
growing an independent branch, is an embarrassing parallel 
problem. Parallelization of the cross-validation is also possible, if each
core tests the Quality function for a different value of the regularization 
parameter $\alpha$.
ROOT libraries are used to handle the 
input-output, but the algorithm is independent, relying on STL containers 
for data structures.

The advantage of DET algorithms over kernel-based density estimators is
the speed of training and evaluation.
The complexity of the algorithm is $\Nbin \times \Ntot$. In common use cases,
the two quantities are not independent, because for larger samples it is 
reasonable to adopt a finer binning in particular in the tails. 
Therefore, depending on the stop condition the computational cost scales with
the size of the data sample as $\Ntot$ to $\Ntot^2$. Kernel density estimation
in the ROOT implementation is found to scale as $\Ntot^2$.

Reading time scales roughly as \Nbin.

\begin{figure}
  \centering
  \includegraphics[width=0.4\textwidth]{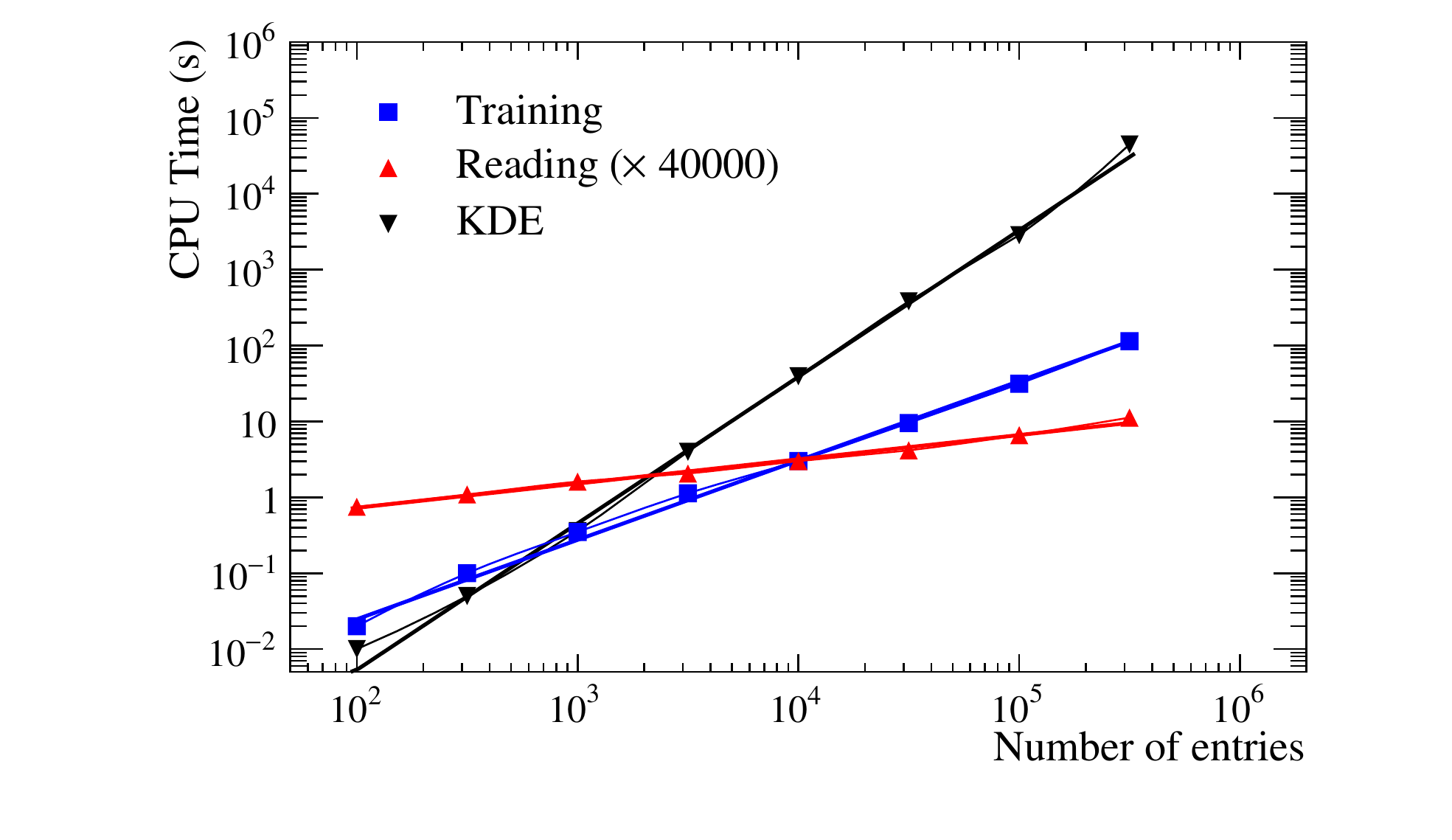}
  \includegraphics[width=0.4\textwidth]{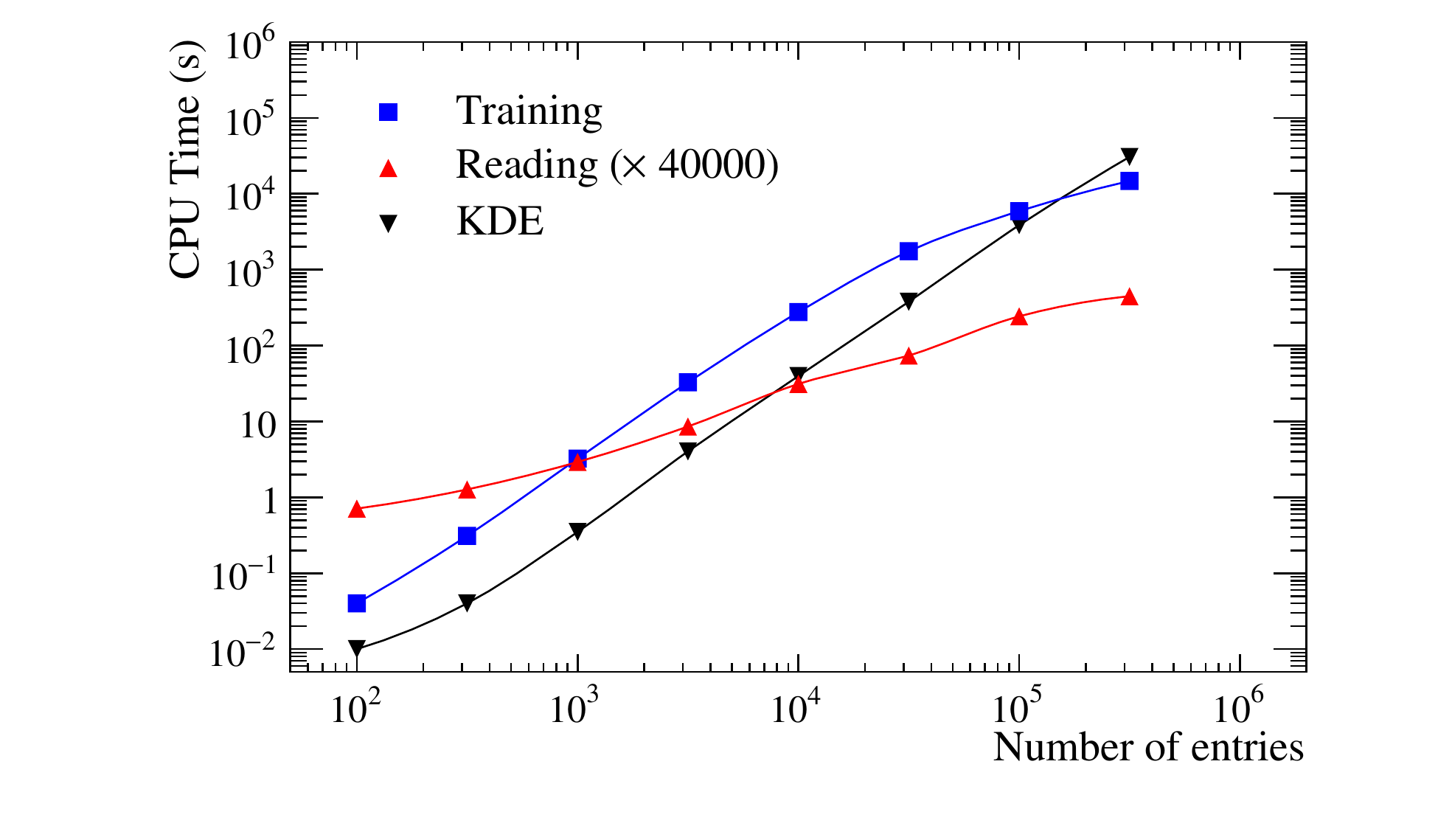}
  \caption{\label{fig:timing}
    CPU time to train and evaluate a self-optimized decision tree as a function
    of the number of entries \Ntot.
    On the top, a stop criterion including a reasonable leaf-width threshold is
    used; on the bottom it is replaced with a very loose threshold. 
    The time needed to train a Kernel Density Estimation (KDE) is also reported
    for comparison.
  }
\end{figure}

Figure \ref{fig:timing} reports the comparison of the CPU time needed to 
train, optimize and sample on a $200\times 200$ grid a DET; the time to train
a kernel density estimation on the same sample is also reported.
The two plots show the results obtained with reasonable and loose stop 
conditions based on the minimal leaf width. It is interesting to observe that 
when using a loose leaf-width condition, $\Nbin \propto \Ntot$ and the 
algorithm scales as $\Ntot^2$. Increasing the size of the sample,
the leaf-width condition becomes relevant and
the computational cost of the DET deflects from $\Ntot^2$, and starts being 
convenient with respect to KDE.

\section{Applications in HEP}
In this section I discuss a few possible use cases of density estimation trees
in High Energy Physics. 
In general, the technique is applicable to all problems involving data modeling,
including efficiency determination and background subtraction. However, for 
these applications KDE is usually preferable, and only in case of too large
samples, in some development phase of the analysis code, it may be reasonable
to adopt DET instead. Here I consider applications where the nature of the 
estimator, providing fast training and fast integration, introduces multivariate 
density estimation into problems traditionally treated alternatively.
The examples are based on a dataset of real data
collected during the $pp$ collision programme of the Large Hadron Collider at
CERN by the LHCb experiment. The dataset has been released by the LHCb 
Collaboration in the framework of the LHCb Masterclass programme. 
The detail of the reconstruction and selection, not relevant to the 
discussion of the DET algorithm, are discussed in Ref. \cite{LHCbMasterClass}.
The data sample contains combinations of a pion ($\pi$) and a 
kaon ($K$), two light mesons,
loosely consistent with the decay of a $D^0$ meson.

\begin{figure}
  \centering
  \includegraphics[width=.5\textwidth]{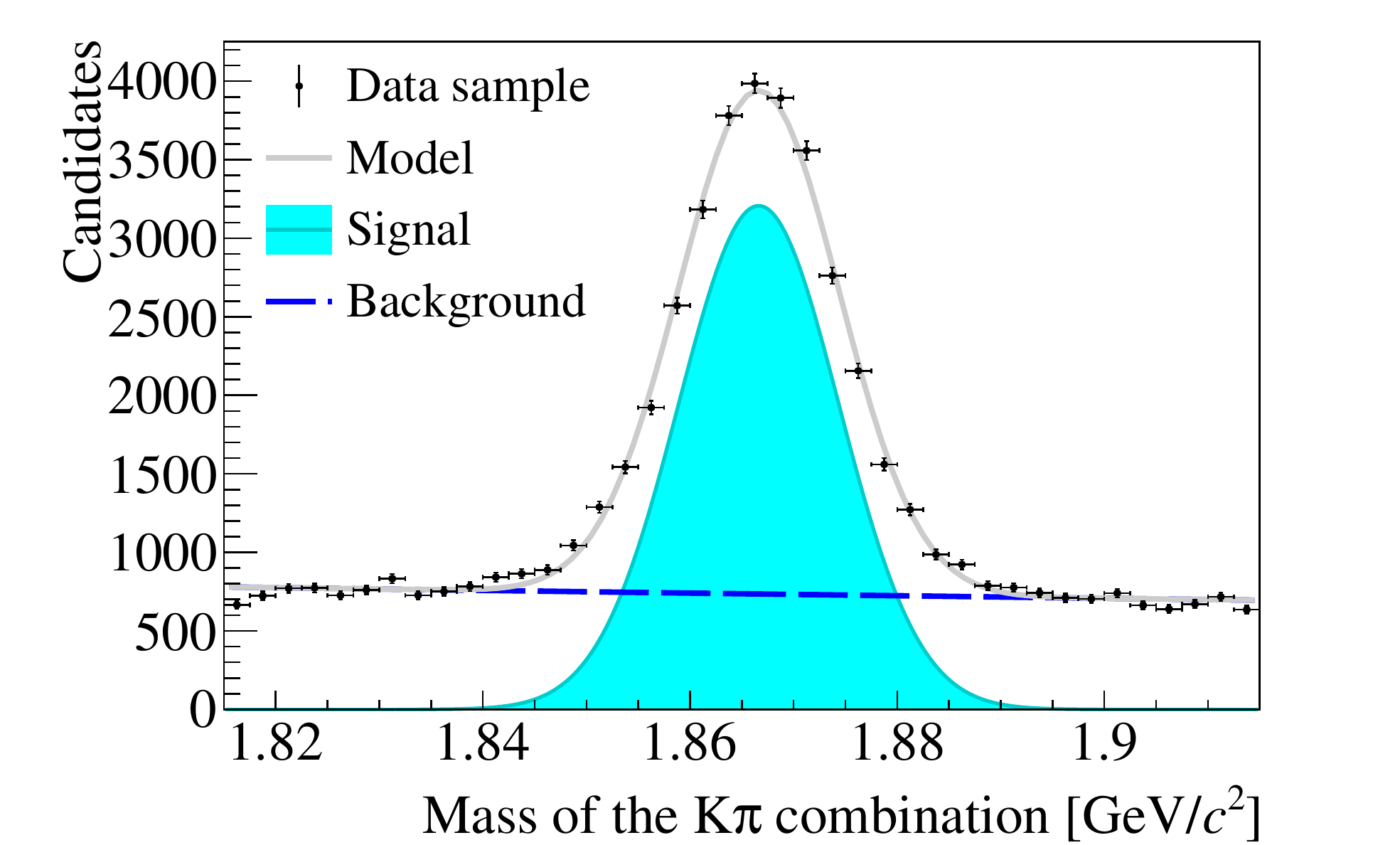}
  \caption{\label{fig:d0mass}
    Invariant mass of the combinations of a kaon and a pion loosely consistent
    with a $D^0$ decay. Two contributions are described in the model: a peaking
    contribution for signal, where the $D^0$ candidates are consistent with
    the mass of the $D^0$ meson (Signal), 
    and a non-peaking contribution due to random
    combinations of a kaon and a pion not produced in a $D^0$ decay 
    (Background).
  }
\end{figure}

Figure \ref{fig:d0mass} shows the invariant mass of the $K\pi$ combination,
\emph{i.e.} the mass of an hypothetical mother particle decayed to the 
reconstructed kaon and pion. Two contributions are evident: a peak due to
real $D^0$ decays, with the invariant mass which is consistent with the 
mass of the $D^0$ meson, and a flat contribution due to the random combination
of kaons and pions, with an invariant mass which is a random number.
The peaked contribution is named ``Signal'', the flat one is the ``Background''.

An important aspect of data analysis in HEP consists in the disentanglement of 
different 
contributions to allow statistical studies of the signal without pollution
from background. In next two Sections, I consider two different approaches 
to signal-background separation. First, an application of DETs to the 
optimization of the rectangular selection is discussed. Then, a more powerful
statistical approach based on likelihood analysis is described.

\subsection{Selection optimization}
When trying to select a relatively pure sample of signal candidates, rejecting 
background, it is important to define an optimal selection strategy based 
on the variables associated to each candidate.
For example, a large momentum of the $D^0$ candidate ($D^0\, p_T$) is more 
common for signal than for background candidates, therefore $D^0$ candidates
with a $p_T$ below a certain threshold can be safely rejected.
The same strategy can be applied to the transverse momentum of the kaon and 
of the pion
separately, which are obviously correlated with the momentum of their mother 
candidate, the $D^0$ meson. Another useful variable is some measure of the 
consistency of the reconstructed flight direction of the $D^0$ candidate with 
its expected origin (the $pp$ vertex). 
Random combinations of a pion and a kaon are likely 
to produce $D^0$ candidates poorly aligned with the point where $D^0$ are 
expected to be produced. In the following I will use the Impact Parameter 
(IP) defined as the distance between the reconstructed 
flight direction of the $D^0$ meson and the $pp$ vertex.

The choice of the thresholds used to reject background to enhance signal
purity often relies on simulated samples of signal candidates, and on 
data regions which are expected to be well dominated by background candidates.
In the example discussed here, the background sample is obtained selecting 
the $D^0$ candidates with a mass $1.815 < m(D^0) < 1.840$ GeV/$c^2$ or 
$1.890 < m(D^0) < 1.915$ GeV/$c^2$.

The usual technique to optimize the selection is to count the number of
simulated signal candidates $N_S$ and background candidates $N_B$ surviving 
a given combination of thresholds $\mathbf{t}$,
and picking the combination which maximizes
some metric $M$, for example 
\begin{equation}
  M(\mathbf{t}) = \frac{S(\mathbf{t})}{S(\mathbf{t})+B(\mathbf{t})+1} 
                = \frac{\epsilon_S N_S(\mathbf{t})}
                {\epsilon_S N_S(\mathbf{t})+\epsilon_B N_B(\mathbf{t})+1} 
\end{equation}
where $\epsilon_S$ ($\epsilon_B$) is the normalization factors between the 
number of entries $N_S^\infty$ ($N_B^\infty$) in the pure sample and 
the expected yields  $S^\infty$ ($B^\infty$) in the mixed sample prior the
selection.

When the number of thresholds to be optimized is large, the optimization may 
require many iterations. Only in absence of correlation between the variables 
used in the selection, the optimization can be factorized reducing the number of 
iterations. For large samples, counting the surviving candidates at each iteration
may become very expensive.

Two DET estimators $\hat f_S(\vecx)$ and $\hat f_B(\vecx)$ for the pure samples 
can be used to reduce the computational
cost of the optimization from \Ntot to \Nbin, integrating the distribution 
leaf by leaf instead of counting the entries.

The integral of the density estimator in the rectangular selection $R$
can be formally written as 
\begin{equation}\label{eq:integral}
  \int_R \hat f(\vecx) \mathrm d\vecx = \frac{1}{\Ntot} \sum_{i = 1}^{\Nbin} 
    \frac{V(\binfunc{i} \cap R)}{\Vbinfunc{i}}\Nbinfunc{i}.
\end{equation}
The optimization requires to find
\begin{equation}
  R = R_{\mathrm{opt}} \quad : \quad M_I(R_{\mathrm{opt}}) = 
    \max_{R \subset \mathcal S} M_I(R),
\end{equation}
with
\begin{equation}
  M_I(R) = \frac{S^{\infty}\int_R \hat f_S(\vecx)\mathrm d \vecx}
                {1 + S^{\infty}\int_R \hat f_S(\vecx)\mathrm d \vecx 
                   + B^{\infty}\int_R \hat f_B(\vecx)\mathrm d \vecx}.
\end{equation}

\begin{figure}
  \centering
  \includegraphics[width=0.4\textwidth]{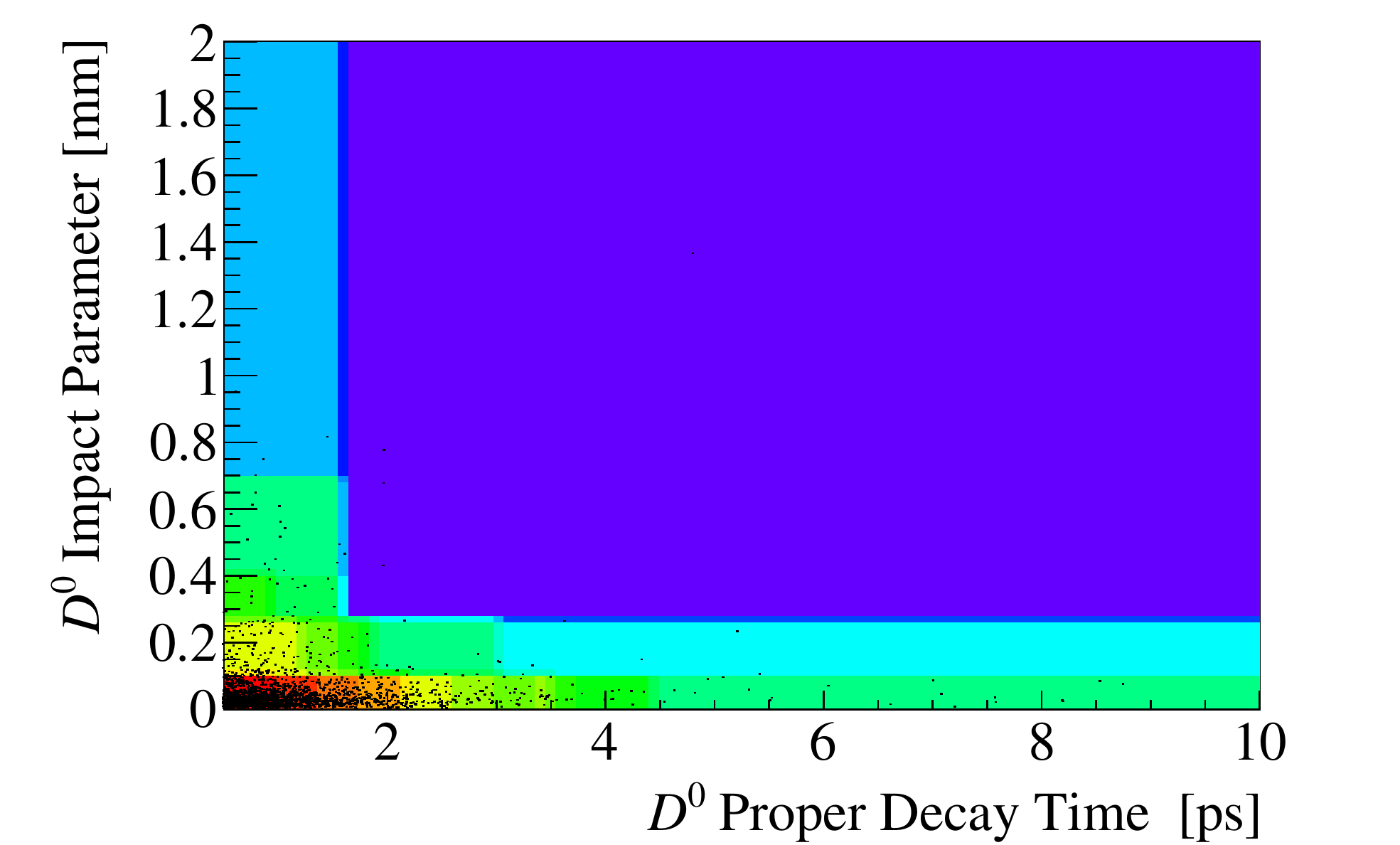}
  \includegraphics[width=0.4\textwidth]{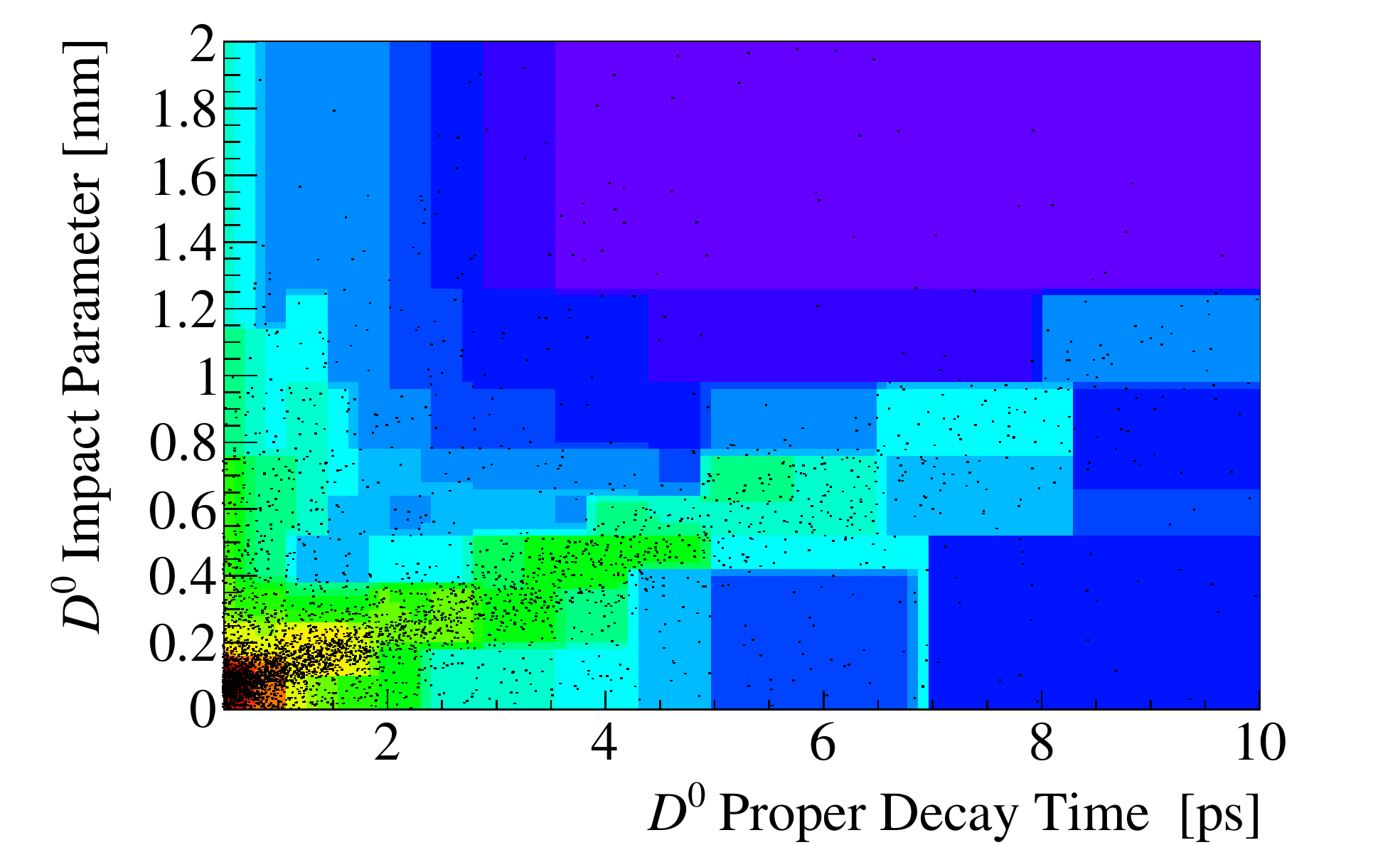}
  \caption{\label{fig:detsigbkg}
    Density Estimation of pure signal (top) and pure background (bottom)
    samples, projected onto the plane of the impact parameter and proper
    decay time. 
    The entries of the data sample are shown as black dots superposed to 
    the color scale representing the density estimation.
  }
\end{figure}

Figure \ref{fig:detsigbkg} reports a projection of $\hat f_S$ and $\hat f_B$ 
onto the plane defined by the impact parameter (IP) and the proper decay time
of the $D^0$ meson. 
The two variables are obviously correlated, because $D^0$ candidates poorly
consistent with their expected origin are associated to a larger decay time
in the reconstruction procedure, which is based on the measurements of the
$D^0$ flight distance and of its momentum. 
The estimation reproduces correctly the correlation, allowing better background 
rejection combining the discriminating power of the two variables when defining
the selection criterion.

\subsection{Likelihood analyses}
Instead of an optimization of the rectangular selection it is reasonable to 
separate signal and background using multivariate techniques as Classification
Trees or Neural Network.

A multivariate statistic based on likelihood can be built using DETs:
\begin{equation}
  \Delta \log\mathcal L (\vecx) = \log\frac{\hat f_S(\vecx)}{\hat f_B(\vecx)}.
\end{equation}

The advantage of using density estimators over Classification Trees is that
likelihood functions from different samples can be easily combined. 
Consider the sample of $K\pi$ combinations described above. Among the 
variables defined to describe each candidate there are Particle Identification 
(PID) variables, response of an Artificial Neural Network (ANN) trained on 
simulation, designed to
provide discrimination, for example, between kaons and pions. The distributions
of PID variables are very difficult to simulate properly because the conditions 
of the detectors used for PID are not perfectly stable during the data 
acquisition. It is therefore preferred to use pure samples of real kaons and 
pions to study the distributions instead of simulating them. 
The distributions obtained depends on the particle momentum $p$, and on the 
angle $\theta$ between the particle momentum and the proton beams.
These variables 
are obviously correlated to the transverse momentum which, as discussed
in the previous section, is a powerful discriminating variable, whose 
distribution has to be taken from simulation, and is in general different 
from simulated samples.
To shorten the equations, 
below I apply the technique to the kaon only, but the same could be done
for the identification of the pion.
The multivariate statistic can therefore be rewritten as 
\begin{align}\nonumber
  \Delta \log & \mathcal L \big(p_T(D^0), \mathrm{IP}, p_T(K), p_T(\pi), p_K, 
      \theta_K, \mathrm{PID}K_K\big) =  \\\nonumber
      = \ & \frac{\hat f_S(p_T(D^0), \mathrm{IP}, p_T(K), p_T(\pi))}
               {\hat f_B(p_T(D^0), \mathrm{IP}, p_T(K), p_T(K),\mathrm{PID}K_K)} 
          \times \\ &  \times 
         \frac{\hat f_K (\mathrm{PID}K_K, p_K, \theta_K)}
               {\int\mathrm d(\mathrm{PID}K_K) 
               \hat f_K (\mathrm{PID}K_K, p_K, \theta_K)},
               \label{eq:mydll}
\end{align}
where PID$K_K$ is the response of the PID ANN for the kaon candidate and the 
kaon hypothesis, and $\hat f_K$ is the DET model built from a pure calibration 
sample of kaons.

The opportunity of operating this disentanglement is due to the properties of 
the probability distribution functions which are not trivially transferable to
Classification Trees. Note that, as opposed to the previous use case, where
integration discourages smearing because
Equation \ref{eq:integral} is not applicable to the smeared version of the 
density estimator, likelihood
analyses can benefit of smearing techniques for the evaluation of the first 
term in Equation \ref{eq:mydll}, while for the second term, smearing can be
avoided thanks to the large statistics usually available for calibration 
samples.

\section{Conclusion}
Density Estimation Trees are fast and robust algorithm providing probability 
density estimators based on decision trees.
They can be grown cheaply beyond overtraining, and then pruned through a
kernel-based cross-validation. The procedure is computationally cheaper than
pure kernel density estimation because the evaluation of the latter is 
performed only once per leaf.

Integration and projections of the density estimator are also fast, providing
an efficient tool for many-variable problems involving large samples.

Smoothing techniques discussed here include smearing and linear interpolation.
The former is useful to fight overtraining, but challenges the adaptability 
of the DET algorithms. Linear interpolation requires tessellation algorithms
which are nowadays available for problems with three or less variables, only.

A few applications to high energy physics have been illustrated using 
the $D^0 \to K^- \pi^+$ decay mode, made public by the LHCb Collaboration
in the framework of the Masterclass programme.
Selection optimization and likelihood analyses can benefit of different
features of the Density Estimation Tree algorithms.
Optimization problems require fast integration of a many-variable density 
estimator, made possible by its simple structure with leaves associated to
constant values.
Likelihood analyses benefit of the speed of the method which allows to 
model large calibration samples in a time much reduced with respect to KDE, 
and offering an accuracy of the statistical model much better than histograms.

In conclusion, Density Estimation Trees are interesting algorithms which can 
play an important role in exploratory data analysis in the field of 
High Energy Physics, filling a gap between the simple histograms and the 
expensive Kernel Density Estimation, and becoming more and more relevant in
the age of the Big Data samples.

\bibliography{det}
\bibliographystyle{icml2015}

\end{document}